# Accurate and Efficient Estimation of Small P-values with the Cross-Entropy Method: Applications in Genomic Data Analysis

Yang Shi[1,2,3,4]*, Mengqiao Wang[2], Weiping Shi[5], Ji-Hyun Lee[3,6], Huining Kang[3,6]* and Hui Jiang[4,7,8]*

[1]Division of Biostatistics and Data Science, Department of Population Health Sciences, Medical College of Georgia, Augusta University, Augusta, Georgia 30912, USA.

[2]West China School of Public Health, Department of Epidemiology and Biostatistics, Sichuan University, Chengdu, Sichuan 610041, P. R. China.

[3]Biostatistics Shared Resource, University of New Mexico Comprehensive Cancer Center and [6]Department of Internal Medicine, University of New Mexico School of Medicine, Albuquerque, New Mexico 87131, USA.

[4]Department of Biostatistics, [7]Center for Computational Medicine and Bioinformatics and [8]University of Michigan Rogel Cancer Center, University of Michigan, Ann Arbor, Michigan 48109, USA.

[5]College of Mathematics, Jilin University, Changchun, Jilin 130012, P. R. China.

*To whom the correspondence should be addressed.

## ABSTRACT

**Motivation:** Small *p*-values are often required to be accurately estimated in large-scale genomic studies for the adjustment of multiple hypothesis tests and the ranking of genomic features based on their statistical significance. For those complicated test statistics whose cumulative distribution functions are analytically intractable, existing methods usually do not work well with small *p*-values due to lack of accuracy or computational restrictions. We propose a general approach for accurately and efficiently estimating small *p*-values for a broad range of complicated test statistics based on the principle of the cross-entropy method and Markov chain Monte Carlo sampling techniques.

**Results:** We evaluate the performance of the proposed algorithm through simulations and demonstrate its application to three real-world examples in genomic studies. The results show that our approach can accurately evaluate small to extremely small *p*-values (e.g. $10^{-6}$ to $10^{-100}$). The proposed algorithm is helpful for the improvement of some existing test procedures and the development of new test procedures in genomic studies.

**Availability:** R programs for implementing the algorithm and reproducing the results are available at: https://github.com/shilab2017/MCMC-CE-codes.

**Contact:** yshi@augusta.edu, hukang@salud.unm.edu and jianghui@umich.edu

**Supplementary information:** Supplementary data are submitted together with the manuscript.

## 1. INTRODUCTION

*P*-value is the most widely used metric to access the statistical significance of genomic features in large-scale genomic studies such as genome-wide association studies (GWAS) and high-throughput differential gene expression analysis. In those studies, very small *p*-values are often required to be accurately calculated, because: (1) A large number of tests are often performed in those studies and most of the methods used for multiple comparison adjustment in genomic studies, such as the Bonferroni correction for family-wise error rate and the Benjamini-Hochberg procedure for controlling false discovery rate (Benjamini and Hochberg, 1995), work directly on the *p*-values associated with the genomic features. Therefore, it is essential to accurately evaluate *p*-values at very small scales so that those procedures are reliable. (2) In practice, it is desirable to rank the significant genomic features by their *p*-values (often together with their effect sizes) so that the researchers can prioritize and follow up with those significant genomic features for further biological studies, which also requires that the small *p*-values associated with those features to be accurately estimated. In the literature, it is not uncommon to see that very small *p*-values associated with the most significant genomic features to the order of less than $10^{-100}$ are reported [e.g. (Burton, et al., 2007) and (Cauchi, et al., 2007). More examples can be found in (Bangalore, et al., 2009)].

### 1.1 Problem formulation

The problem addressed in this work is how to estimate small *p*-values for a group of complicated test statistics whose cumulative distribution functions (CDF) are analytically intractable. Specifically, the question can be formulated as follows: the goal is to estimate the *p*-value defined as

$$p = Pr[T(\mathbf{Y}) \geq q \mid H_0], \qquad (1)$$

where $\mathbf{Y}$ is the data or transformed data that follow some probability distribution (e.g. multivariate normal distribution), $T(\mathbf{Y})$ is the test statistic which is a function of $\mathbf{Y}$, $q$ is the test statistic calculated based on the observed data that can be either a scalar or a vector, and $H_0$ means that the probability is obtained under the null hypothesis, which will be dropped for simplicity hereafter. In most commonly used test procedures (e.g. two-sample *t*-test), the *p*-values are obtained by deriving the exact or asymptotic distributions of $T(\mathbf{Y})$ under $H_0$. However, the problem we often encounter is that $T(\mathbf{Y})$ is complicated whose CDF under $H_0$ cannot be derived analytically, and existing approaches do not work for estimating very small *p*-values either due to lack of accuracy or unaffordable computational burden.

We illustrate this problem with the following three real-world examples in genomic studies.

***Example 1: Gene set/pathway enrichment analysis***.

Here the goal is to test the significance of the association between some clinical outcomes of interest and the global expression pattern of a gene set or pathway (for brevity, gene set will be used hereafter), where gene sets are pre-specified groups of genes according to the biological functions or genomic locations of the genes. For a study with $n$ independent subjects and a gene set with $k$ genes, Goeman *et al* proposed to fit the following model,

$$E(\mathbf{y}) = g^{-1}(\mathbf{X}\boldsymbol{\alpha} + \mathbf{Z}\boldsymbol{\beta}), \quad (2)$$

where $\mathbf{y}$ is the $n \times 1$ outcome vector, $\mathbf{Z}$ is an $n \times k$ expression matrix of the $k$ genes in $n$ subjects, $\mathbf{X}$ is an $n \times p$ matrix for the $p$ covariates that needs to be adjusted, $g$ is the canonical link function for the distribution of $\mathbf{y}$ (e.g. $g$ is the identity function for normally distributed data or the logit function for binomially distributed data), and $\boldsymbol{\alpha}$ and $\boldsymbol{\beta}$ are the corresponding vectors of coefficients. The association between $\mathbf{Z}$ and outcome $\mathbf{y}$ can be assessed by testing the null hypothesis $H_0 : \boldsymbol{\beta} = \mathbf{0}$ using the following test statistic

$$Q_1 = n^{-1}(\mathbf{y}-\boldsymbol{\mu})^T \mathbf{Z}\mathbf{Z}^T(\mathbf{y}-\boldsymbol{\mu}), \tag{3}$$

where $\boldsymbol{\mu}$ is the expectation of $\mathbf{y}$ under $H_0$ (Goeman, et al., 2004; Goeman, et al., 2011). A similar approach is also proposed in (Liu, et al., 2007), where the matrix $\mathbf{Z}\mathbf{Z}^T$ in Eq. (3) is replaced by a kernel function to account for the interaction of genes in the same gene set.

***Example 2: GWAS – joint testing a group of genetic markers in a genomic region***

To increase the power of GWAS, approaches for joint testing a group of genetic markers (SNPs) in a genomic region instead of testing individual genetic markers are developed. Wu *et al* proposed an approach under similar framework as in Example 1, and the following model is fit

$$E(\mathbf{y}) = g^{-1}(\mathbf{X}\boldsymbol{\alpha} + \mathbf{G}\boldsymbol{\beta}), \tag{4}$$

where $\mathbf{y}$ is the $n \times 1$ phenotype vector, $\mathbf{G}$ is an $n \times s$ genotype matrix for the $s$ SNPs in the genomic region that need to be tested, $\mathbf{X}$ is an $n \times p$ matrix for the $p$ covariates that needs to be adjusted, $g$ is the canonical link function for the distribution of $\mathbf{y}$, and $\boldsymbol{\alpha}$ and $\boldsymbol{\beta}$ are the vectors of coefficients. The association between $\mathbf{G}$ and phenotype $\mathbf{y}$ can be assessed by testing the null hypothesis $H_0 : \boldsymbol{\beta} = \mathbf{0}$ using the following test statistic

$$Q_2 = (\mathbf{y}-\boldsymbol{\mu})^T \mathbf{G}\mathbf{W}\mathbf{G}^T(\mathbf{y}-\boldsymbol{\mu}), \tag{5}$$

where $\boldsymbol{\mu}$ is the expectation of $\mathbf{y}$ under $H_0$ and $\mathbf{W}$ is a diagonal matrix containing the weights of the $s$ SNPs (Wu, et al., 2011). If the weight of each SNP is 1, then $Q_2$ is the same as $Q_1$ in Example 1 up to a constant. Similarly, the matrix $\mathbf{G}\mathbf{W}\mathbf{G}^T$ can be replaced by a kernel matrix to account for the epistatic effects of the SNPs (Wu, et al., 2011).

***Example 3: Ratio statistics in differential gene expression analysis***

Consider the differential expression analysis comparing two groups of gene expression data. For a gene $g$ to be tested, let $\mathbf{x}_{g1}$ and $\mathbf{x}_{g2}$ be the vectors of positive normalized gene expression values respectively for the two groups with sample sizes $n_1$ and $n_2$. The following ratio statistic (also known as fold change or proportion statistic) has been proposed to test the differential expression of $g$ between the two groups (Segal, et al., 2017),

$$L = \max(y_1 / y_2, y_2 / y_1), \tag{6}$$

where $y_1$ and $y_2$ are the respective sample means of the two groups. Note that the $p$-value computed based on (6) is the two-sided $p$-value based on the test statistic, $y_1 / y_2$ (Segal, et al., 2017). Without loss of generality and for the ease of derivation, we assume $y_1 \geq y_2$ and use the test statistic,

$$L = y_1 / y_2, \tag{7}$$

in the following discussions. Other approaches for testing differential gene expression based on the ratio statistics are also proposed in (Bergemann and Wilson, 2011; Chen, et al., 2002; Newton, et al., 2001).

The exact or asymptotic $p$-values with the test statistics in the above three examples can be expressed as the general form (1). To see this, for Examples 1 and 2, it can be shown that the test statistics $Q_1$ and $Q_2$ can be written in the following quadratic form

$$Q = T(\mathbf{Y}) = \mathbf{Y}^T \mathbf{D} \mathbf{Y}, \tag{8}$$

where $\mathbf{Y}$ follows a multivariate normal (MVN) distribution either exactly normal if the outcome or phenotype data $\mathbf{y}$ is assumed to follow normal distributions or asymptotically normal if $\mathbf{y}$ is assumed to follow binomial distributions, and $\mathbf{D}$ is a diagonal matrix containing the positive eigenvalues of $\mathbf{Z}\mathbf{Z}^T$ in $Q_1$ or $\mathbf{G}\mathbf{W}\mathbf{G}^T$ in $Q_2$. See (Duchesne and De Micheaux, 2010) for matrix calculations about how $Q_1$ and $Q_2$ can be expressed as $Q$. Therefore, the $p$-values with the test statistics $Q_1$ and $Q_2$ can be expressed in the form $Pr[T(\mathbf{Y}) \geq q_1]$, where $q_1$ is the test statistic calculated based on the observed data. For Example 3, let $q_2$ be the ratio statistic calculated based on the observed data, then the $p$-value with the ratio statistic $L$ can be further expressed as

$$p = Pr(y_1 / y_2 \geq q_2 \mid y_1, y_2 > 0) = Pr(\mathbf{CY} > \mathbf{0}), \tag{9}$$

where $\mathbf{C} = \begin{pmatrix} 0 & 1 \\ 1 & -q_2 \end{pmatrix}$ and $\mathbf{Y} = (y_1, y_2)^T$. Based on the central limit theorem, $y_1$ and $y_2$ respectively follow $N(\mu, \sigma^2 / n_1)$ and $N(\mu, \sigma^2 / n_2)$ either exactly if the expression data are assumed to be normal or asymptotically if the expression data are not normal, where $\mu$ and $\sigma^2$ are the population mean and variance, respectively, under the null hypothesis that there is no differential expression between the two groups.

### 1.2 Related works

In the literature, the quadratic form statistics (8), also known as a linear combination or weighted sum of chi-squared random variables (Bausch, 2013), are used for testing the associations between genomic features and the outcomes or phenotypes under several settings in genomic studies. A few methods are proposed specifically to calculate the tail probabilities for this form of statistic (Davies, 1980; Farebrother, 1984; Imhof, 1961; Liu, et al., 2009). See (Duchesne and De Micheaux, 2010) for comparisons of them and (Bausch, 2013) for a review. As commented in (Bausch, 2013), most of those existing methods do not work well when $p$ is very small due to lack of accuracy or computational restrictions. In addition, those approaches are aimed specifically for the quadratic form $Q$ and cannot be generalized to other types of test statistics, which have a relative narrow scope of applications.

Alternative to the above methods, a more general type of approaches is the utilization of Monte Carlo (MC) sampling methods, where a large number of MC random samples can be generated under $H_0$ either via simulations from the distribution of $\mathbf{Y}$ under $H_0$ (Lin, 2005) or through permutation or bootstrap of the observed data (note that permutation and bootstrap are special cases of MC methods, where the former samples the observed data without replacement and the latter samples with replacement), and then repeatedly calculate the test statistics using those MC samples and estimate $p$ as the proportion of the test statistics based on the MC samples that are greater than or equal to the one based on the observed data. Yet if we want to accurately estimate very small $p$-values, this type of brute-force MC methods will need enormous computational effort.

In this paper, we propose a general approach for accurately and efficiently estimating small to extremely small $p$-values for any test statistic that can be expressed in the form (1). The basis of our approach contains two components. The first component is the cross-entropy (CE) method, which is originated from the concept of CE in information theory and has been widely used for rare event simulation in the operations research field (Rubinstein and Kroese, 2004). The second component is Markov chain Monte Carlo (MCMC) sampling techniques. Therefore, we refer our approach as MCMC-CE algorithm hereafter. The rest of this paper is organized as follows. In the next section, we give a general introduction of the CE method and MCMC techniques used in our approach, and then present our algorithm for estimating small $p$-values. In Section 3, we evaluate the performance of the proposed algorithm by comparing it with several existing approaches through simulations and demonstrate its applications with three real genomic datasets. Discussions are given in Section 4.

## 2. METHODS

### 2.1 The CE method

Our goal is to calculate the $p$-value as expressed in Eq. (1), which can be further written as

$$p = Pr[T(\mathbf{Y}) \geq q] = E_{\boldsymbol{\theta}_0}[I\{T(\mathbf{Y}) \geq q\}], \tag{10}$$

where the subscript $\boldsymbol{\theta}_0$ denotes the parameter vector of the probability distribution that $\mathbf{Y}$ follows under $H_0$ [e.g. it is an MVN distribution in the above three examples. We use $f(\cdot;\boldsymbol{\theta}_0)$ to denote this distribution hereafter], and the expectation is taken with respect to $f(\cdot;\boldsymbol{\theta}_0)$ with $I(\cdot)$ as the indicator function.

As discussed above, when $p$ is small, using the brute-force MC method is computationally inefficient. The CE method is a general approach for the efficient estimation of small probabilities in MC simulations, which we briefly introduce below following the monograph on CE method (Rubinstein and Kroese, 2004). The technique used in the CE method is importance sampling (IS). Let $g(\cdot)$ be the proposal density function used in IS, then the expectation in Eq. (10) can be re-expressed as

$$p = \int I\{T(\mathbf{Y}) \geq q\} \frac{f(\mathbf{Y};\boldsymbol{\theta}_0)}{g(\mathbf{Y})} g(\mathbf{Y}) \, \mathrm{d}\mathbf{Y} = E_g[I\{T(\mathbf{Y}) \geq q\} \frac{f(\mathbf{Y};\boldsymbol{\theta}_0)}{g(\mathbf{Y})}], \tag{11}$$

where the subscript $g$ denotes that the expectation is taken with respect to $g(\cdot)$ now. Then $p$ can be estimated by the MC counterpart (also known as stochastic counterpart) of (11),

$$\hat{p} = \frac{1}{N} \sum_{l=1}^{N} I\{T(\mathbf{y}_l) \geq q\} \frac{f(\mathbf{y}_l;\boldsymbol{\theta}_0)}{g(\mathbf{y}_l)}, \tag{12}$$

where $\mathbf{y}_l$'s, $l$ = 1, ..., $N$ are random samples drawn from $g(\cdot)$ now. There is an optimal proposal density under which the IS estimator (12) has zero MC sampling variance (Rubinstein and Kroese, 2004), which is given by

$$g^*(\mathbf{Y}) = \frac{I\{T(\mathbf{Y}) \geq q\} f(\mathbf{Y};\boldsymbol{\theta}_0)}{p}. \tag{13}$$

However, $g^*$ cannot be directly used as the proposal density for estimating $p$ in IS, since it contains the unknown probability $p$ that is the quantity we want to calculate. The CE method provides a general solution to finding a proposal density $f(\cdot;\boldsymbol{\theta})$ which is close to the optimal proposal density $g^*$ within the same distribution family as $f(\cdot;\boldsymbol{\theta}_0)$ in the sense that the Kullback–Leibler divergence [also known as the Kullback–Leibler cross-entropy or cross-entropy (Rubinstein and Kroese, 2004)] between $f(\cdot;\boldsymbol{\theta})$ and $g^*$ is minimized:

$$\begin{aligned} D\left(g^*(\cdot), f\left(\cdot;\boldsymbol{\theta}\right)\right) &:= \int g^*(\mathbf{Y}) \ln \frac{g^*(\mathbf{Y})}{f\left(\mathbf{Y};\boldsymbol{\theta}\right)} \mathrm{d}\,\mathbf{y} \\ &= \int g^*(\mathbf{Y}) \ln g^*(\mathbf{Y}) \,\mathrm{d}\,\mathbf{Y} - \int g^*(\mathbf{Y}) \ln f\left(\mathbf{Y};\boldsymbol{\theta}\right) \mathrm{d}\,\mathbf{Y} \end{aligned}, \tag{14}$$

where $f(\cdot;\boldsymbol{\theta})$ is another distribution within the same family as $f(\cdot;\boldsymbol{\theta}_0)$, and we will call $f(\cdot;\boldsymbol{\theta})$ the CE-optimal proposal density below. Since the first term on the right-hand side of the second equality in Eq. (14) does not contain $\boldsymbol{\theta}$, the parameter $\boldsymbol{\theta}$ that minimizes $D\left(g^*(\cdot), f\left(\cdot;\boldsymbol{\theta}\right)\right)$ should maximize the second term. Hence, the problem of finding the CE-optimal proposal density $f(\cdot;\boldsymbol{\theta})$ turns into an optimization problem of finding $\boldsymbol{\theta}$ that maximizes the second term, $\int g^*(\mathbf{Y}) \ln f\left(\mathbf{Y};\boldsymbol{\theta}\right) \mathrm{d}\,\mathbf{Y}$. Originally, Rubinstein *et al* developed an adaptive algorithm to solve this optimization problem, which is referred as "multi-level

CE method" in the literature (Rubinstein and Kroese, 2004). We review that algorithm and discuss its limitations in detail in Supplementary Text Section 1.

One of the major limitations of the multi-level CE algorithm is that it is unreliable in high-dimensional settings (i.e. when the dimension of $\boldsymbol{\theta}$ is large). With recent progress in MCMC sampling techniques, here we apply and implement an improved version of CE method based on the theoretical work in (Chan and Kroese, 2012) that combines the CE criterion (14) and MCMC techniques. Observe that the second term on the right-hand side of the second equality in (14) can be written as

$$\int g^*(\mathbf{Y}) \ln f\left(\mathbf{Y};\boldsymbol{\theta}\right) \mathrm{d}\,\mathbf{Y} = E_{g^*}[\ln f\left(\mathbf{Y};\boldsymbol{\theta}\right)], \qquad (15)$$

where the subscript $g^*$ means that the expectation is taken with respect to the optimal proposal density $g^*$. Therefore, if we can draw random samples from $g^*$ (in the next section, we will discuss the techniques to that end), the parameter $\boldsymbol{\theta}$ that minimizes $D\left(g^*(\cdot), f\left(\cdot;\boldsymbol{\theta}\right)\right)$ can be found by the maximization of the expectation in (15). By replacing the expectation in (15) with its MC counterpart, $\boldsymbol{\theta}$ can be found by solving

$$\arg\max_{\boldsymbol{\theta}} \frac{1}{N} \sum_{l=1}^{N} [\ln f\left(\mathbf{y}_l;\boldsymbol{\theta}\right)], \qquad (16)$$

where $\mathbf{y}_l$'s, $l$ = 1, ..., $N$ are random samples drawn from $g^*$. Problem (16) is a regular maximum likelihood estimation problem, which can be solved either analytically or numerically by widely used approaches such as Newton-Raphson or the Expectation-Maximization algorithms.

## 2.2 Sampling from the optimal proposal density

Here we discuss the algorithms for sampling from $g^*$. From (13), observe that $g^*$ is a truncated distribution with $f(\cdot;\boldsymbol{\theta}_0)$ restricted by the constraint $T(\mathbf{Y}) \geq q$. With the recent progress in MCMC techniques, several algorithms are developed for sampling from a truncated distribution in the form of $g^*$ and Table 1 summarizes four of them. The Gibbs sampler is a classical MCMC method for sampling from truncated distributions that consecutively draws samples from a sequence of conditional distributions (Geweke, 1991; Kotecha and Djuric, 1999). The hit-and-run sampler belongs to the class of line samplers, which reduces the problem of sampling from a multivariate constrained distribution to that of sampling from a univariate truncated distribution (Chen and Schmeiser, 1993; Kroese, et al., 2011). The Hamiltonian and Lagrangian Monte Carlo samplers are two more recently developed powerful tools for sampling from many complicated distributions, which respectively use the principles of the Hamiltonian and Lagrangian dynamics in physics (Lan, et al., 2015; Pakman and Paninski, 2014). In our empirical comparisons, we find that the Hamiltonian and Lagrangian Monte Carlo samplers are more efficient for sampling from a truncated distribution with quadratic constraints such as Example 1 and 2, while the Gibbs sampler is more efficient for sampling from a truncated distribution with linear constraints such as Example 3 (not shown here). More comparisons of those algorithms can be found in (Kroese, et al., 2011; Lan, et al., 2015; Pakman and Paninski, 2014).

**Table 1.** Algorithms for sampling from the optimal proposal distribution

| **Algorithm** | **Reference** |
|---|---|
| Gibbs sampler | Geweke, 1991; Kotecha and Djuric, 1999 |
| Hit-and-run sampler | Chen and Schmeiser, 1993; Kroese, et al., 2011 |
| Hamiltonian Monte Carlo sampler | Brubaker, et al., 2012; Pakman and Paninski, 2014 |
| Lagrangian Monte Carlo sampler | Lan, et al., 2015 |

### 2.3 The MCMC-CE algorithm for calculating small *p*-values

Combining the above discussions, our algorithm for estimating the small $p$-value $p = Pr[T(\mathbf{Y}) \geq q]$ contains two steps: in the first step we draw random samples from $g^*$ and solve the maximization problem (16) to obtain the CE-optimal proposal density $f(\cdot;\boldsymbol{\theta})$, and in the second step we estimate $p$ using standard IS with $f(\cdot;\boldsymbol{\theta})$ as the proposal density. The algorithm is summarized as follows:

**Main Algorithm (**MCMC-CE method for estimating small $p$-values**)**

A. Parameter updating step:

1. Draw $N$ random samples $\mathbf{y}_1,...,\mathbf{y}_N$ from $g^*$ using an efficient MCMC sampling algorithm (as shown in Table 1).

2. Solve the maximization problem (16) and obtain the CE-optimal proposal density $f(\cdot;\boldsymbol{\theta})$.

B. Estimating step:

Draw $M$ random samples $\mathbf{y}_1,...,\mathbf{y}_M$ from $f(\cdot;\boldsymbol{\theta})$. Estimate $p$ as

$$\hat{p} = \frac{1}{M}\sum_{l=1}^{M} I\{T(\mathbf{y}_l) \geq q\}\frac{f(\mathbf{y}_l;\boldsymbol{\theta}_0)}{f(\mathbf{y}_l;\boldsymbol{\theta})}.$$

## 3. RESULTS

### 3.1 Simulation studies

We perform simulations to evaluate the performance of the MCMC-CE algorithm. In the first part of the simulations, we investigate the estimation accuracy and computational efficiency of MCMC-CE via simulations from random variables whose tail probabilities can be calculated analytically. Our simulations focus on calculating $Pr[T(\mathbf{Y}) \geq q]$, where $\mathbf{Y}$ is an MVN random variable. Specifically, we use the following two types of random variables whose true tail probabilities are available in most statistical packages, so that we can evaluate the errors and variations of MCMC-CE: (1) chi-squared random variables, which can be expressed as a special case of quadratic functions of MVN random variables, and (2) standard Cauchy random variable, which is the ratio of two independent standard normal variables and can be expressed as a special case of linear functions of MVN random variables. The details of the simulations are given in Supplementary Text Section 2.1 and the results are shown in Supplementary Tables S1-S5, which are briefly summarized below.

In the first experiment, we use the following four $\chi_m^2$'s (denotes a chi-squared random variable with $m$ degrees of freedom): $\chi_5^2$, $\chi_{20}^2$, $\chi_{50}^2$ and $\chi_{100}^2$, where the dimensions increase from 5 to 100. With each of them, we generate a sequence of true $p$-values on the order from $10^{-6}$ to $10^{-100}$ and compare the results of MCMC-CE and several other methods specific for calculating the tail probabilities of the quadratic form statistic (8), including Davies' method (Davies, 1980), Farebrother's method (Farebrother, 1984) and Imhof's method (Imhof, 1961). The details of this experiment are given in Supplementary Text Section 2.1 and the results are shown in Supplementary Tables S1 - S4. This experiment shows that MCMC-CE can accurately calculate $p$ to the order of $10^{-100}$ with less than 5% relative errors in all the four settings, while none of other methods in the comparisons works well when $p$ is smaller than $10^{-16}$. Figure 1 shows a graphical demonstration of the concordance between the $p$-values estimated using MCMC-CE and the true $p$-values for chi-squared random variables with different degrees of freedom.

In the second experiment, we use MCMC-CE to calculate the small tail probabilities of the standard Cauchy random variable and compare the results with the true values. The details of this experiment are given in Supplementary Text Section 2.1 and the results are shown in Supplementary Table S5. The experiment shows that MCMC-CE can accurately calculate *p*-values to the order of $10^{-100}$ with less than 3% relative errors.

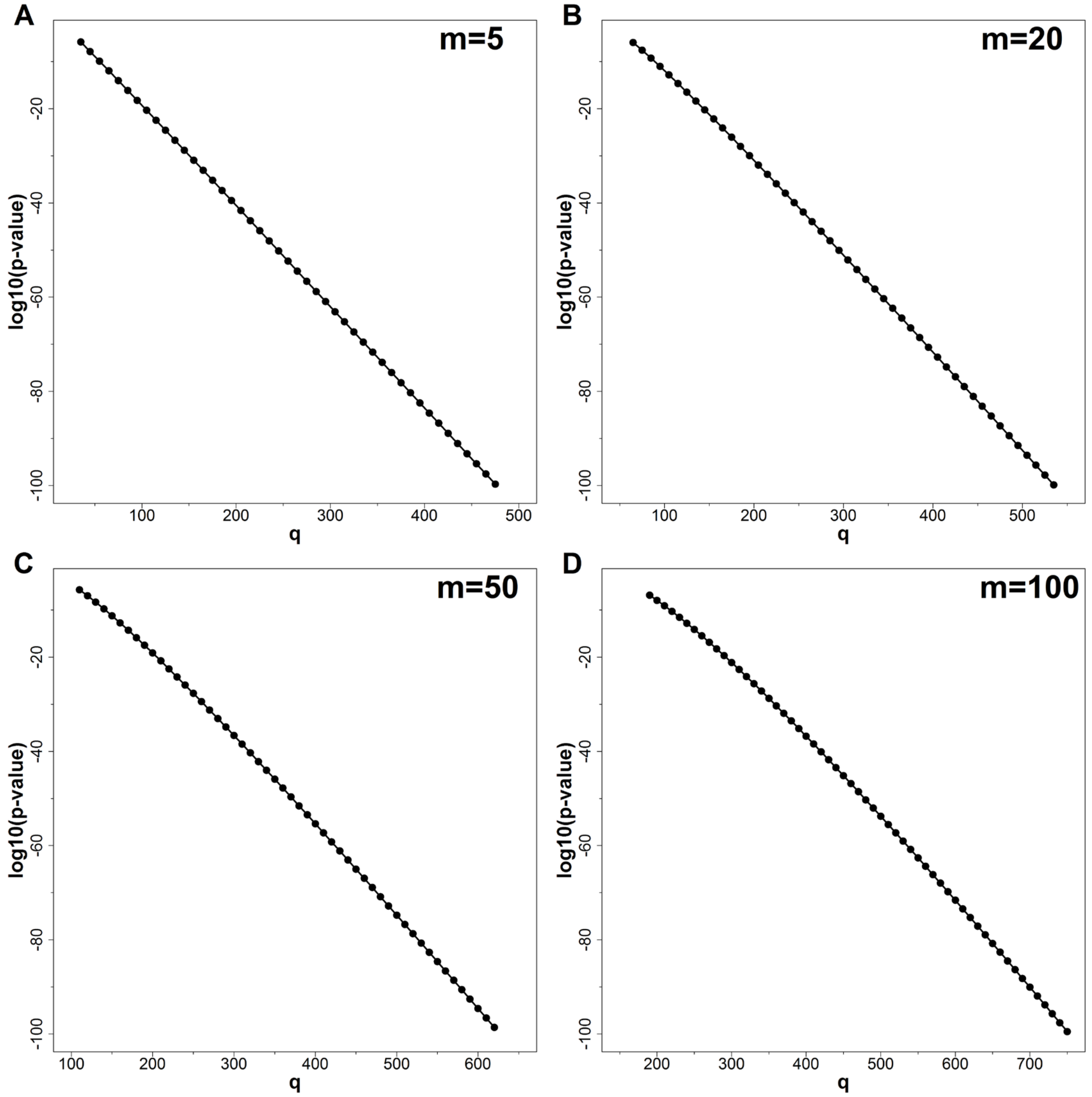

**Fig. 1. Concordance between the true tail probabilities of the chi-squared variables and the results from MCMC-CE.** In each figure panel, the solid line represents the true tail probabilities and the dots represent the ones estimated from MCMC-CE. The detailed results are presented in Supplementary Tables S1 – S4.

In the second part of the simulation studies, we perform sensitivity analysis to examine the estimation accuracy and computational efficiency of MCMC-CE with different numbers of random samples in the parameter updating step and estimating steps (i.e. *N* and *M* in the Main Algorithm), where we use MCMC-CE to estimate the *p*-values on the orders from $10^{-6}$ to $10^{-100}$ based on four chi-squared random variables, $\chi_5^2$, $\chi_{20}^2$, $\chi_{50}^2$ and $\chi_{100}^2$ with combinations of different *N*s and *M*s. The details are given in Supplementary Text Section 2.2 and the results are presented in Supplementary Table S6. The results show that the variations of MCMC-CE (assessed using the metric standardized root-mean-square error, *SRMSE*) are reduced by increasing *N* and *M*. When *N* grows large enough with a fixed *M*, the variations no longer decrease, which indicates that the parameter updating step is stabilized. Once the parameter updating step is stabilized with a large enough *N*, the *SRMSE* decreases roughly proportionally with the square root of *M*. The variations can be well-controlled (less than 5%) with large enough *N* and *M* even when the dimensionality of the parameter space is very large and the *p*-value is extremely small, and the computation time is affordable on a typical desktop computer (Supplementary Table S6).

### 3.2 Application to genomic data analysis

We apply MCMC-CE to the estimation of small *p*-values in three real-world examples from genomic studies.

***Example 1: Gene set/pathway enrichment analysis***.

We apply MCMC-CE to a dataset containing gene expression measurements and clinical variables of melanoma patients, which is part of The Cancer Genome Atlas (TCGA) project and publicly available from TCGA data portal: https://portal.gdc.cancer.gov/.

Particularly, the dataset contains the expression levels of 20531 genes from 355 melanoma patients measured by RNA-Seq, and we are interested in testing which gene sets are associated with the clinical variable of interest, Breslow thickness. The gene set annotations are extracted from the Gene Ontology Consortium (Ashburner, et al., 2000; Gene Ontology Consortium, 2017), where 22211 gene sets are curated. For each gene set, model (2) is fitted with log transformed Breslow thickness as the outcome variable and gender and age as the adjusted covariates. Since the computational time will be overwhelming if all the 22211 gene sets are to be tested using MCMC-CE, we first use the following screening test to filter out those less-significant gene sets: we calculate the approximated $p$-values for all the gene sets using the method implementing in the *globaltest* package (Goeman, et al., 2004), where the distribution of the test statistic $Q_1$ as defined in Eq. (3) is approximated by a scaled chi-square distribution. Based on this screening test, there are 35 gene sets with the approximated $p$-values less than $10^{-8}$, and we use MCMC-CE to accurately calculate the $p$-values associated with those 35 gene sets. For each gene set, the algorithm is run 100 times to obtain the variations of the estimated $p$-value, and $N$ = 10000 random samples in the parameter updating step and $M$ = 10000 random samples in the estimating step are used. Supplementary Table S7 presents the results for all the 35 gene sets and Table 2 presents the top 10 most significant ones. The results show that MCMC-CE can efficiently calculate extremely small $p$-values to the order of $10^{-54}$ (Table 2). For comparison, two other approaches for calculating the $p$-values for the quadratic form statistic (8), the Davies' and Farebrother's methods, are also applied in this example. Similar to the observations in the simulations, neither of those two methods works when the $p$-values are smaller than $10^{-16}$, and they agree with the results from MCMC-CE in the cases where the $p$-values are not too small (Supplementary Table S7).

**Table 2.** Top ten gene sets significantly associated with Breslow thickness ranked by their *p*-values.

| GO term | No. gene | *P*-value | S.D. | Time (Average) | Time (SD) | Description |
|---|---|---|---|---|---|---|
| GO:0048880 | 5 | $2.29\times10^{-54}$ | $4.73\times10^{-56}$ | $2.45\times10^{-1}$ | $2.28\times10^{-2}$ | sensory system development |
| GO:1900019 | 5 | $7.17\times10^{-53}$ | $1.10\times10^{-54}$ | $2.58\times10^{-1}$ | $2.36\times10^{-2}$ | regulation of protein kinase C activity |
| GO:1900020 | 5 | $7.20\times10^{-53}$ | $1.67\times10^{-54}$ | $2.63\times10^{-1}$ | $2.12\times10^{-2}$ | positive regulation of protein kinase C activity |
| GO:0045499 | 9 | $1.25\times10^{-48}$ | $2.12\times10^{-50}$ | $2.30\times10^{-1}$ | $1.99\times10^{-2}$ | chemorepellent activity |
| GO:0004415 | 8 | $1.50\times10^{-45}$ | $3.09\times10^{-47}$ | $2.54\times10^{-1}$ | $2.30\times10^{-2}$ | hyalurononglucosaminidase activity |
| GO:0016941 | 3 | $5.05\times10^{-45}$ | $7.49\times10^{-47}$ | $2.37\times10^{-1}$ | $2.41\times10^{-2}$ | natriuretic peptide receptor activity |
| GO:0007168 | 10 | $5.19\times10^{-45}$ | $9.91\times10^{-47}$ | $2.43\times10^{-1}$ | $2.13\times10^{-2}$ | receptor guanylyl cyclase signaling pathway |
| GO:2000020 | 8 | $6.46\times10^{-35}$ | $2.08\times10^{-36}$ | $1.97\times10^{-1}$ | $1.71\times10^{-2}$ | positive regulation of male gonad development |
| GO:2000018 | 9 | $7.21\times10^{-35}$ | $2.14\times10^{-36}$ | $1.87\times10^{-1}$ | $1.75\times10^{-2}$ | regulation of male gonad development |
| GO:0045163 | 3 | $1.85\times10^{-34}$ | $3.61\times10^{-36}$ | $2.26\times10^{-1}$ | $1.98\times10^{-2}$ | clustering of voltage-gated potassium channels |

**No. gene**: number of genes in the gene set; ***P*-value**: *p*-value estimated from MCMC-CE, which is the average of the results of 100 runs of the algorithm; **S.D.**: standard deviation of the results of 100 runs. **Time (Average)** and **Time (SD)**: the average and standard deviation of the real elapsed time in seconds for a single run of MCMC-CE on a single core of Intel Xeon X5550 2.67GHz CPU calculated based on 100 runs of the algorithm. See Table S7 for the detailed results.

***Example 2: GWAS – joint testing a group of genetic markers in a genomic region***

We demonstrate the application of MCMC-CE in testing groups of SNPs in GWAS. The dataset used is collected in a GWAS performed in the population of about 2000 heterogeneous stock mice phenotyped for over 100 traits (Valdar, et al., 2006), which is public available at: https://wp.cs.ucl.ac.uk/outbredmice/heterogeneous-stock-mice/. We are interested in testing which regions of SNPs are associated with the serum concentration of high-density lipoprotein (HDL). After preprocessing, the dataset contains 1640 subjects with complete HDL data and 10990 SNPs with complete genotype data. For demonstration purpose, we simply group each adjacent 20 SNPs based on their genomic locations as one region, which results in 549 groups of SNPs. For each group of SNPs, model (4) is fitted with HDL as the phenotype variable and gender and weight of the mice as the adjusted covariates, and MCMC-CE is used to calculate the $p$-value with the test statistic (5). Similar to Example 1, the algorithm is run 100 times to obtain the variations of the estimated $p$-value, and $N$ = 10000 and $M$ = 10000 random samples are used. Supplementary Table S8 presents the 70 groups of SNPs that are significantly associated HDL with $p$-values less than $10^{-8}$ and Table 3 shows the top 10 most significant ones. Those results demonstrate that MCMC-CE can efficiently calculate extremely small $p$-values to the order of $10^{-38}$. Our analyses based on SNP groups are also in agreement with the tests for individual SNP reported in the original study (Valdar, et al., 2006). We also applied the Davies' and Farebrother's methods in this example, and same as before, neither of them works when the $p$-values are smaller than $10^{-16}$ (Supplementary Table S8).

**Table 3.** Top ten groups of SNPs significantly associated with HDL ranked by their $p$-values

| SNP group index | ***P*-value** | **S.D.** | **Time (Average)** | **Time (SD)** |
|---|---|---|---|---|
| Group 42 | $5.93\times10^{-38}$ | $1.11\times10^{-39}$ | $2.00\times10^{-1}$ | $1.78\times10^{-2}$ |
| Group 44 | $5.65\times10^{-36}$ | $3.80\times10^{-37}$ | $2.28\times10^{-1}$ | $1.91\times10^{-2}$ |
| Group 200 | $1.51\times10^{-19}$ | $6.27\times10^{-21}$ | $1.76\times10^{-1}$ | $1.59\times10^{-2}$ |
| Group 214 | $1.88\times10^{-19}$ | $1.78\times10^{-20}$ | $2.43\times10^{-1}$ | $1.81\times10^{-2}$ |
| Group 43 | $6.06\times10^{-19}$ | $1.98\times10^{-20}$ | $1.66\times10^{-1}$ | $1.93\times10^{-2}$ |
| Group 314 | $1.03\times10^{-17}$ | $7.38\times10^{-18}$ | $1.89\times10^{-1}$ | $1.77\times10^{-2}$ |
| Group 528 | $7.61\times10^{-17}$ | $8.03\times10^{-18}$ | $2.06\times10^{-1}$ | $1.66\times10^{-2}$ |
| Group 213 | $4.64\times10^{-16}$ | $9.71\times10^{-18}$ | $1.84\times10^{-1}$ | $1.65\times10^{-2}$ |
| Group 276 | $6.09\times10^{-16}$ | $9.56\times10^{-17}$ | $2.12\times10^{-1}$ | $1.74\times10^{-2}$ |
| Group 273 | $7.71\times10^{-16}$ | $1.13\times10^{-17}$ | $1.83\times10^{-1}$ | $1.87\times10^{-2}$ |

The meanings of ***P*-value**, **S.D.**, **Time (Average)** and **Time (SD)** are the same as Table 2. See Table S8 for the detailed results.

*Example 3: Ratio statistic in differential gene expression analysis*

We demonstrate the application of MCMC-CE in estimating small $p$-values based on the ratio statistic (6) in differential gene expression analysis and show how MCMC-CE can be used to assess the genome-wide significance after the adjustment of multiple comparisons. The dataset used is from a study on patients who were diagnosed with salivary adenoid cystic carcinoma and received radiation therapy, and the expression levels of 22243 genes in the salivary gland tissues of those patients were measured by RNA-Seq. The details of the study can be found in (Brayer, et al., 2016) and the sequencing read data are available from the NCBI Sequence Read Archive using accession number SRP059557. The dataset used consists of 14 patients, where 8 were free of cancer and 6 were not at the end of the study, and here we are interested in testing genes differentially expressed between those two groups of patients. After filtering out lowly expressed genes, 11390 genes are left and the gene count data are normalized using the trimmed mean of M-values method implemented in R package *edgeR* (Robinson, et al., 2010) and log-transformed counts per million (CPM) values are used for our analysis.

The following methods for estimating the $p$-values based on the ratio statistic are included in our comparisons:

(1) A brute-force MC method: for each gene, the two-sided $p$-value is computed as $\hat{p} = 2\times\min[\frac{1}{M_{BF}}\sum_{l=1}^{M} I(\frac{y_{1l}}{y_{2l}} \geq q), \frac{1}{M_{BF}}\sum_{l=1}^{M} I(\frac{y_{1l}}{y_{2l}} \leq q)]$, where $y_{1l}$'s and $y_{2l}$'s, $l = 1, ..., M_{BF}$ are MC samples drawn from the normal distributions described in Section 1.1, $q$ is the statistic computed based on the observed data, and $M_{BF}$, the number of MC samples, is set as $10^5$.

(2) MCMC-CE: based on the results of the above brute-force MC method, there are 150 genes with $p$-values less than $10^{-4}$. We use MCMC-CE to accurately compute the $p$-values for those 150 genes, then we combine the results with those genes with $p$-values greater than $10^{-4}$ as the final results. Here $N$ = 10000 random samples in the parameter updating step and $M$ = 10000 random samples in the estimating step are used, and the algorithm is run 100 times for each $p$-value to assess the variations.

(3) A permutation method: for each gene, the $p$-value is computed as $\hat{p} = \frac{1}{N_{perm}} \sum_{i=1}^{N} I(L_i \geq q)$, where $L_i$'s, $i$ = 1, ..., $N_{perm}$ are the test statistic as defined in (6) computed based on the permutations of the observed gene expression data and $q$ is the same statistic computed based on the observed data. Here, the set of all permutations of the observed data can be enumerated (i.e. $N$ = 3003). We also use an alternative formula to compute the $p$-value as $\hat{p} = \frac{1}{N_{perm}+1} [\sum_{i=1}^{N} I(L_i \geq q) + 1]$. The two formulas are respectively referred as "Perm0" and "Perm1" below.

For all the above methods, we use the Benjamini–Hochberg procedure (Benjamini and Hochberg, 1995) to control the FDR given the $p$-values computed by each method. As a comparison, we also run the differential expression analysis using *samr* package with its default settings (Tusher, et al., 2001). Table 4 presents the numbers of significantly differentially expressed genes identified by each method with different FDR threshold values, and Supplemental Table S8 shows the detailed results. We can see that the brute-force MC method suffers the issue that the results with more stringent FDR thresholds are not reliable (Table 4, FDR threshold = 0.01, 0.005 and 0.001), as the small *p*-values with those most significant genes cannot be accurately estimated due to the limited number of MC samples. Those permutation-based approaches (Perm0, Perm1 and *samr*) suffer the same issue, though it should be noted that the null hypotheses tested by them are different from the brute-force MC and MCMC-CE. This example shows that accurate estimation of small *p*-values is useful for correctly evaluating the genome-wide significance. Of note, the smallest *p*-value estimated by MCMC-CE in this example is $1.11\times10^{-15}$ (Supplementary Table S8). We also illustrate the application of MCMC-CE for differential expression analysis in a microarray dataset with larger sample size and more extreme *p*-values, which is given in Supplementary Text Section 3.

**Table 4.** Number of significantly differentially expressed genes identified by each method with different FDR thresholds

| | **FDR threshold** | | | | | |
|---|---|---|---|---|---|---|
| | **0.001** | **0.005** | **0.01** | **0.05** | **0.1** | **0.15** |
| **Brute-force MC** | 52 | 52 | 52 | 105 | 195 | 296 |
| **MCMC-CE** | 17 | 30 | 40 | 82 | 190 | 296 |
| **Perm0** | 31 | 31 | 31 | 31 | 31 | 57 |
| **Perm1** | 0 | 0 | 0 | 0 | 0 | 0 |
| **samr** | 7 | 7 | 7 | 22 | 56 | 226 |

## 4. DISCUSSION

In summary, we propose an algorithm, MCMC-CE for accurate and efficient estimation of small *p*-values for test statistics that can be expressed in the form of Eq. (1) and demonstrate its application in genomic data analyses. To apply MCMC-CE, the following requirements should be met: (1) The test statistic needs to be written as a function of $\mathbf{Y}$, and $\mathbf{Y}$ follows a certain distribution that belongs to a parametric family of distributions, as denoted by $f(\cdot;\cdot)$ in Section 2; (2) It is feasible to generate random samples from $f(\cdot;\cdot)$; (3) The density of $f(\cdot;\cdot)$ can be evaluated pointwisely. Nonetheless, MCMC-CE can estimate small $p$-values for a broad range of test statistics. Although we demonstrate the application of MCMC-CE for the test statistics that are quadratic or linear functions of MVN random variables in the simulations and real-world examples in this paper, MCMC-CE can also work for more complicated test statistics that are not limited to the functions of MVN random variables (Chan and Kroese, 2012; Pakman and Paninski, 2013). Therefore, it can help researchers develop new test procedures in genomic studies.

Since both the parameter updating and estimating steps of MCMC-CE involve drawing random samples and the simulations also show that the variations of the algorithm tend to increase when the dimensionality of the parameter space grows and the $p$-values become smaller (Supplementary Text Section 2.1), here we discuss the sources of variations and provide some strategy for the choice of the numbers of random samples in the parameter updating step (i.e. $N$) and in the estimating step (i.e. $M$) for the practical usage of the algorithm based on the sensitivity analysis (Supplementary Text Section 2.2 and Supplementary Table S6). Basically, the variations are from the following sources:

(1) The variations of random sampling in the parameter updating step, which can affect the estimation of the CE-optimal proposal density in Eq. (16). This portion of variations can be reduced by increasing $N$. When $N$ is large enough with a fixed $M$, the overall variations no longer decrease, which indicates that the parameter updating step is stabilized (Supplementary Text Section 2.2 and Supplementary Table S6).

(2) The variations of random sampling from the CE-optimal proposal density in the estimating step. This portion of variations can be reduced by increasing $M$. As shown in the simulations, once the parameter updating step is stabilized with large enough $N$, the standardized root-mean-square error decreases roughly proportionally with the square root of $M$ (Supplementary Table S6). This agrees with the general Monte Carlo method, where the Monte Carlo sampling error decreases proportionally with the square root of the number of random samples (Kroese, et al., 2011).

(3) Since MCMC-CE uses the CE-optimal proposal density $f(\cdot;\boldsymbol{\theta})$ to approximate the optimal proposal density $g^*$, $f(\cdot;\boldsymbol{\theta})$ still does not behave exactly like $g^*$, which leads to the intrinsic variations of this method. This portion of variations gains its weight with the "curse of dimensionality" (i.e. when the dimensionality of the parameter space grows), which can be seen by comparing the variations in the simulation results from $\chi_5^2$ to $\chi_{100}^2$ (Supplementary Table S1 - S4).

In practice, sensitivity analysis like Supplementary Table S6 is recommended for the choice of optimal $N$ and $M$ when applying MCMC-CE. A general strategy is to first find the optimal $N$ by examining the variations across different values of $N$ with a fixed $M$. Once the optimal $N$ is determined, we can then increase $M$ to achieve a desired level of accuracy.

## Acknowledgements

We thank Drs. Maureen Sartor and Xiaoquan Wen at University of Michigan for reading and helpful discussions on Section 2, which is part of YS's doctoral dissertation (Shi, 2016). The computational resources used in this work were provided by the University of New Mexico Center for Advanced Research Computing and Augusta University Medical College of Georgia, Department of Population Health Sciences.

## Funding

This work was partly supported by the following grants: one startup research grant from Augusta University Medical College of Georgia (YS), two startup research grants from Sichuan University supported by the Fundamental Research Funds for the Central Universities of China (20822041B4009 of YS and 20822041A4202 of MW), the National Natural Science Foundation of China Grant J1310022 (WS), NIH grant 5P30CA118100 (YS, JL and HK) and NIH grant P30CA046592 (HJ).

*Conflict of Interest:* none declared.

# Accurate and Efficient Estimation of Small P-values with the Cross-Entropy Method: Applications in Genomic Data Analysis

## Supplementary Text

Yang Shi[1,2,3,4]*, Mengqiao Wang[2], Weiping Shi[5], Ji-Hyun Lee[3,6], Huining Kang[3,6]* and Hui Jiang[4,7,8]*

[1]Division of Biostatistics and Data Science, Department of Population Health Sciences, Medical College of Georgia, Augusta University, Augusta, Georgia 30912, USA.

[2]West China School of Public Health, Department of Epidemiology and Biostatistics, Sichuan University, Chengdu, Sichuan 610041, P. R. China.

[3]Biostatistics Shared Resource, University of New Mexico Comprehensive Cancer Center and [6]Department of Internal Medicine, University of New Mexico School of Medicine, Albuquerque, New Mexico 87131, USA.

[4]Department of Biostatistics, [7]Center for Computational Medicine and Bioinformatics and [8]University of Michigan Rogel Cancer Center, University of Michigan, Ann Arbor, Michigan 48109, USA.

[5]College of Mathematics, Jilin University, Changchun, Jilin 130012, P. R. China.

*To whom the correspondence should be addressed. Email: yshi@augusta.edu, hukang@salud.unm.edu and jianghui@umich.edu

## 1. Review of the multi-level CE method

Here we briefly review the multi-level CE method, which is an earlier alternative to the MCMC-CE method presented in the main text. Our goal is to find $\boldsymbol{\theta}$ that maximize the second term on the right-hand side of the second equality in Eq. (14) in the main text, $\int g^*(\mathbf{Y})\ln f\left(\mathbf{Y};\boldsymbol{\theta}\right)\mathrm{d}\,\mathbf{Y}$. The multi-level CE method uses Eq. (13) in the main text and expresses the second term on the right-hand side of the second equality in Eq. (14) in the main text as

$$\begin{aligned}\int g^*(\mathbf{Y})\ln f\left(\mathbf{Y};\boldsymbol{\theta}\right)\mathrm{d}\,\mathbf{Y} &= \int \frac{I\{T(\mathbf{Y})\geq q\}f(\mathbf{Y};\boldsymbol{\theta}_0)}{p}\ln f\left(\mathbf{Y};\boldsymbol{\theta}\right)\mathrm{d}\,\mathbf{Y} \\ &= \frac{1}{p}E_{\boldsymbol{\theta}_0}[I\{T(\mathbf{Y})\geq q\}\ln f\left(\mathbf{Y};\boldsymbol{\theta}\right)]\end{aligned}. \quad \text{(S1)}$$

Therefore, $\boldsymbol{\theta}$ should maximize (S1), which is the solution to the following problem:

$$\arg\max_{\boldsymbol{\theta}} E_{\boldsymbol{\theta}_0}[I\{T(\mathbf{Y})\geq q\}\ln f\left(\mathbf{Y};\boldsymbol{\theta}\right)]. \quad \text{(S2)}$$

The key idea of the multi-level CE method [Chapter 3 of (Rubinstein and Kroese, 2004)] is to solve problem (S2) adaptively via importance sampling (IS). By using IS again with the proposal density $f(\cdot;\boldsymbol{\theta}_k)$, problem (S2) can be expressed as

$$\arg\max_{\boldsymbol{\theta}} E_{\boldsymbol{\theta}_k}[I\{T(\mathbf{Y})\geq q\}\frac{f(\mathbf{Y};\boldsymbol{\theta}_0)}{f(\mathbf{Y};\boldsymbol{\theta}_k)}\ln f(\mathbf{Y};\boldsymbol{\theta})]. \quad \text{(S3)}$$

The MC counterpart of (S3) is

$$\arg\max_{\boldsymbol{\theta}} \frac{1}{N}\sum_{l=1}^{N}[I\{T(\mathbf{y}_l)\geq q\}\frac{f(\mathbf{y}_l;\boldsymbol{\theta}_0)}{f(\mathbf{y}_l;\boldsymbol{\theta}_k)}\ln f(\mathbf{y}_l;\boldsymbol{\theta})], \quad \text{(S4)}$$

where $\mathbf{y}_l$'s, $l = 1,\dots,N$ are random samples drawn from the proposal density $f(\cdot;\boldsymbol{\theta}_k)$.

Rubinstein *et al* developed the following multi-level algorithm for solving problem (S4):

**Algorithm S1** (multi-level CE algorithm for estimating small probabilities)

A. Adaptive updating step:

(1) Specify a constant $\rho \in (0,1)$. Start with parameter $\boldsymbol{\theta}_0$; Set the iteration counter $k = 0$.

(2) At the $k$th iteration, generate random samples $\mathbf{y}_1,...,\mathbf{y}_N$ from $f(\cdot;\boldsymbol{\theta}_k)$. Calculate the statistics $T(\mathbf{y}_1),\ldots,T(\mathbf{y}_N)$, and compute $q_k$ as their sample $(1-\rho)$ quantile provided $q_k < q$. If $q_k \geq q$, set $q_k = q$.

(3) Updating the parameter $\boldsymbol{\theta}_k$ with $\boldsymbol{\theta}_{k+1}$, which is the solution to the following problem:

$$\boldsymbol{\theta}_{k+1} = \arg\max_{\boldsymbol{\theta}} \frac{1}{N}\sum_{l=1}^{N}[I\{T(\mathbf{y}_l) \geq q_k\}\frac{f(\mathbf{y}_l;\boldsymbol{\theta}_0)}{f(\mathbf{y}_l;\boldsymbol{\theta}_k)}\ln f(\mathbf{y}_l;\boldsymbol{\theta})]. \quad \text{(S5)}$$

Note that the difference between (S4) and (S5) is $q$ in (S4) is substituted by $q_k$ in (S5).

(4) If $q_k < q$, set $k = k + 1$ and reiterate from Step (2); else, proceed to the following step.

B. Estimating step:

Use $f(\cdot;\boldsymbol{\theta}_k)$ as the proposal density and generate random samples $\mathbf{y}_1,...,\mathbf{y}_M$ from it. Estimate $p$ as $\hat{p} = \frac{1}{M}\sum_{l=1}^{M}[I\{T(\mathbf{y}_l) \geq q\}\frac{f(\mathbf{y}_l;\boldsymbol{\theta}_0)}{f(\mathbf{y}_l;\boldsymbol{\theta}_k)}]$.

Here we briefly discuss the rationale of the above multi-level CE algorithm: The adaptive updating step of the algorithm iteratively generates a sequence of updated parameters $\{\boldsymbol{\theta}_k, k = 0,1...\}$ and a sequence of threshold values $\{q_k, k = 0,1...\}$. According to the work of Rubinstein *et al*, under mild regularity conditions, $\{q_k, k = 0,1...\}$ is monotonically non-decreasing and the target threshold value $q$ can be reached with high probability in a finite number of iterations for small $\rho$ (Rubinstein, 1999; Rubinstein and Kroese, 2004). Hence, the updated parameters $\{\boldsymbol{\theta}_k, k = 0,1...\}$ is more and more close to the optimal parameter $\boldsymbol{\theta}$ that we want to find in problem (S4). The estimating step is simply a standard importance sampling that uses $f(\cdot;\boldsymbol{\theta}_k)$ as the proposal density.

Based on the relevant literature and our studies of implementing and testing the above multi-level CE algorithm (Shi, 2016; Shi, et al., 2016), we find this algorithm have the following two limitations: First, it becomes unreliable with large biases and variations in high-dimensional settings, i.e. when the dimension of $\boldsymbol{\theta}_0$ is high. As discussed in (Chan and Kroese, 2012; Rubinstein and Glynn, 2009), one important reason is that the likelihood ratio $\frac{f(\mathbf{y}_l;\boldsymbol{\theta}_0)}{f(\mathbf{y}_l;\boldsymbol{\theta}_k)}$ involved in (S5) becomes unstable in high dimensions, thus the proposal density $f(\cdot;\boldsymbol{\theta}_k)$ obtained from Algorithm S1 can be far from the optimal proposal density $g^*$. Second, the determination of the parameter $\rho$ is heuristic, which often needs trial and error. Therefore, we apply the improved version of CE method (Chan and Kroese, 2012) in estimating small $p$-values as discussed in Section 2.1 in the main text. The improved CE method can overcome the above limitations of the multi-level CE method but requires sampling from the optimal proposal density $g^*$ as discussed in Section 2.2 of the main text.

## 2. Simulation studies

We perform simulations to evaluate the performance of the MCMC-CE algorithm. Our simulation studies contain two parts. In the first part, we investigate the accuracy and computation efficiency of MCMC-CE via simulations from random variables whose tail probabilities can be calculated analytically. In the second part, we examine the accuracy and efficiency of MCMC-CE with different numbers of random samples in the parameter updating step (i.e. $N$ in the Main Algorithm in the main text) and in the estimating step (i.e. $M$ in the Main Algorithm in the main text).

## 2.1 Study the accuracy and efficiency of MCMC-CE

Here we apply MCMC-CE for the estimation of small tail probabilities of random variables with known cumulative distribution functions, so that those small tail probabilities can be calculated analytically and we can access the errors and variations of MCMC-CE. Our simulations below focus on calculating $Pr[T(\mathbf{Y}) \geq q]$, where $\mathbf{Y}$ is a multivariate normal (MVN) random variable. We performed two numerical experiments: In the first experiment, we use MCMC-CE to estimate the small tail probabilities of the chi-squared random variables, which can be expressed as a special case of quadratic form of MVN variables. In the second experiment, we use MCMC-CE to estimate the small tail probabilities of standard Cauchy random variable, which is the ratio of two independent standard normal random variables and can be expressed as a special case of linear functions of MVN random variables.

Before presenting the simulation studies, we briefly discuss the MCMC algorithms for sampling from the optimal proposal density $g^*$ as listed in Table 1 of the main text. The Gibbs sampler is a classical MCMC method for sampling from truncated distributions that consecutively draws samples from a sequence of conditional distributions (Geweke, 1991; Kotecha and Djuric, 1999). The hit-and-run sampler belongs to the class of line samplers, which reduces the problem of sampling from a multivariate constrained distribution to that of sampling from a univariate truncated distribution (Chen and Schmeiser, 1993; Kroese, et al., 2011). The Hamiltonian and Lagrangian Monte Carlo samplers are two more recently developed powerful tools for sampling from many complicated distributions, which respectively use the principles of the Hamiltonian and Lagrangian dynamics in physics (Lan, et al., 2015; Pakman and Paninski, 2014). In our empirical comparisons, we find that the Hamiltonian and Lagrangian Monte Carlo samplers are more efficient for sampling from a truncated distribution with quadratic constraints, while the Gibbs sampler is more efficient for sampling from a truncated distribution with linear constraints. Therefore, we use the Hamiltonian Monte Carlo sampler when applying MCMC-CE to the simulations with the chi-squared random variables in the first experiment, and the Gibbs sampler to the simulations with the standard Cauchy random variable in the second experiment.

***Experiment 1: simulations with chi-squared random variables***

Here we use MCMC-CE to evaluate the small tail probabilities of quadratic form of MVN random variables and compare it with several approaches specific for calculating the tail probabilities of this form of statistic, including Davies' method (Davies, 1980), Farebrother's method (Farebrother, 1984), and Imhof's method (Imhof, 1961), which are all implemented in the R package *CompQuadForm* (Duchesne and De Micheaux, 2010).

Our simulations are based on the chi-squared random variables. Note that a chi-squared random variable $\chi_m^2$ with $m$ degrees of freedom can be written as

$$\chi_m^2 = \mathbf{Y}^T \mathbf{D} \mathbf{Y} ,$$

where $\mathbf{Y}$ follows $\text{MVN}(\mathbf{0}_{m\times 1}, \mathbf{I}_{m\times m})$ and $\mathbf{D} = \text{diag}(1,...,1)$ is a diagonal matrix with $m$ 1's on the diagonal. Therefore, $\chi_m^2$ is a special case of quadratic form of MVN random variables whose true tail probabilities are available in most statistical packages (e.g. it can be calculated from the *pchisq* function in R), so that we can evaluate the errors of MCMC-CE and other approaches.

We use the following four $\chi_m^2$'s: $\chi_5^2$, $\chi_{20}^2$, $\chi_{50}^2$ and $\chi_{100}^2$ in the experiment. For each of them, we use a sequence of $q$'s and obtain the $p$-values $p = Pr(\mathbf{Y}^T \mathbf{D} \mathbf{Y} \geq q)$ on orders ranging from $10^{-6}$ to $10^{-100}$, and we use different approaches (Davies, Farebrother, Imhof and MCMC-CE) to estimate $p$'s and the errors. For MCMC-CE, we generate $N = 10^4$ random samples in the parameters updating step using the Hamiltonian Monte Carlo sampler (Table 1 in the main text) and $M = 10^4$ random samples in the estimating step (see the Main Algorithm in the main text). Furthermore, to assess the variations of MCMC-CE, we repeat the algorithm 100 times for the estimation of each $p$, and use the average of $\hat{p}$'s from the 100 runs of MCMC-CE as the final estimation of $p$. For each method, the absolute value of relative error (*ARE*) is used to assess the error, which is defined as

$$ARE = \left| \frac{\hat{p} - p}{p} \right|, \tag{S6}$$

where $\hat{p}$ is the estimated $p$-value by each method and $p$ is the true tail probability. To assess the variations of MCMC-CE, a new metric, the standardized root-mean-square error (*SRMSE*) is used, which is defined as

$$SRMSE = \frac{\sqrt{\sum_{i=1}^{N} (\hat{p}_i - p)^2 / n}}{p}, \tag{S7}$$

where $n$ is the number of runs of the algorithm (i.e. 100 in this example), $\hat{p}_i$ is the estimated $p$ of the $i$th run and $p$ is the true tail probability.

The simulation results together with the computation time of each method are presented in Supplementary Tables S1 – S4. We can see that Davies' method works well when $p$ is on the orders up to $10^{-9}$~$10^{-10}$; Farebrother's method works well when $p$ is on the orders up to $10^{-15}$~$10^{-16}$; Imhof's method works well when $p$ is on the order of $10^{-9}$~$10^{-10}$. None of those methods works well when $p$ is smaller than $10^{-16}$. In all the four settings, MCMC-CE can accurately calculate $p$ up to the order of $10^{-100}$ with less than 5% relative errors (Supplementary Table S1-S4, Column **ARE_MCMC-CE**). For variations of MCMC-CE (Supplementary Table S1-S4, Column **SRMSE_MCMC-CE**), we can see that *SRMSE* is controlled at less than 15% for all the $p$-values in low-dimensional settings ( $\chi_5^2$ in Supplementary Table S1 and $\chi_{20}^2$ in Supplementary Table S2) and for $p$-values to the order of $10^{-80}$ with $\chi_{50}^2$ (Supplementary Table S3) or $10^{-60}$ with $\chi_{100}^2$ (Supplementary Table S4), while it increases to 20% - 30% for extremely small $p$-values (~$10^{-100}$) in high-dimensional settings ( $\chi_{50}^2$ in Supplementary Table S3 and $\chi_{100}^2$ in Supplementary Table S4). To reduce the variations for those extremely small $p$-values in high-dimensional settings, we need to increase the numbers of random samples in both the parameter updating step and the estimating step of MCMC-CE, and we further study this issue in the following Section 2.2. Regarding the computation time of MCMC-CE, it grows with the increment of the dimensionality of the parameter space (from $\chi_5^2$ to $\chi_{100}^2$) and the decrement of the $p$-values (from $10^{-6}$ to $10^{-100}$, Supplementary Table S1 – S4), but it is affordable on typical desktop computers.

***Experiment 2: simulations with standard Cauchy random variables***

Here we use MCMC-CE to calculate the small tail probabilities of the ratio of two independent normal random variables, $p = Pr(y_1 / y_2 \geq q)$, where $y_1$ follows $N(\mu_1, \sigma_1^2)$, $y_2$ follows $N(\mu_2, \sigma_2^2)$ and $q$ is a scalar. This ratio can be written as a linear function of MVN random variables (see also Example 3 in the main text):

$$p = Pr(\mathbf{C}_1\mathbf{Y} > \mathbf{0}) + Pr(\mathbf{C}_2\mathbf{Y} > \mathbf{0}),$$

where $\mathbf{C}_1 = \begin{pmatrix} 0 & 1 \\ 1 & -q \end{pmatrix}$, $\mathbf{C}_2 = \begin{pmatrix} 0 & -1 \\ -1 & q \end{pmatrix}$ and $\mathbf{Y} = \left(y_1, y_2\right)^T$ following MVN[$\begin{pmatrix} \mu_1 \\ \mu_2 \end{pmatrix}$, $\begin{pmatrix} \sigma_2^2 & 0 \\ 0 & \sigma_2^2 \end{pmatrix}$]. Note that we do not add the constraint $y_1, y_2 > 0$ here as in Example 3 in the main text.

Specifically, the ratio of two independent standard normal variables follows the standard Cauchy distribution, whose true tail probabilities are available in most statistical packages (e.g. it can be calculated from the *pcauchy* function in R). Therefore, our simulations here are based on the ratio of two independent standard normal random variables so that we can evaluate the errors and variations of MCMC-CE. With the standard Cauchy distribution, we use a sequence of $q$'s and obtain the values of $p$ on orders ranging from $10^{-6}$ to $10^{-100}$, and we use MCMC-CE to calculate $p$'s and the errors. Here we generate $N = 10^4$ random samples in the parameter updating step using the Gibbs sampler (Table 1 in the main text) and $M = 10^4$ random samples in the estimating step (see the Main Algorithm in the main text). Similar to the previous experiment, we repeat the algorithm 100 times for the estimation of each $p$, and use the average of $\hat{p}$'s from the 100 runs of MCMC-CE as the final estimation of $p$. Similarly, we use *ARE* and *SRMSE* as defined in Eqs. (S6) and (S7) to access the errors and variations of MCMC-CE. The results are presented in Supplementary Table S5. We can see that MCMC-CE can accurately estimate $p$ to the order of $10^{-100}$ with less than 3% relative errors and less than 10% *SRMSE* (Supplementary Table S5).

**2.2 Sensitivity analysis with different numbers of random samples**

Since both the parameter updating and estimating steps of MCMC-CE involve drawing random samples and the above simulations show that the variations of the algorithm tend to increase when the dimensionality of the parameter space grows (from $\chi_5^2$ to $\chi_{100}^2$) and the *p*-value becomes smaller (from $10^{-6}$ to $10^{-100}$), here we perform sensitivity analysis to examine the estimation accuracy and computational efficiency of MCMC-CE with different numbers of random samples in the parameter updating step (i.e. *N* in the Main Algorithm in the main text) and in the estimating step (i.e. *M* in the Main Algorithm in the main text). Similar to the first experiment in Section 2.1, we use MCMC-CE to estimate the *p*-values on the orders from $10^{-6}$ to $10^{-100}$ based on four chi-squared distributions, $\chi_5^2$, $\chi_{20}^2$, $\chi_{50}^2$ and $\chi_{100}^2$, but with various combinations of *N*s and *M*s. Supplementary Table S6 presents the results, where the algorithm is run 100 times for each combination of *N* and *M*, and the *SRMSE* as defined in Eq. (S7) and the computation time are calculated and shown (Supplementary Table S6). The results show that the variations of the algorithm are reduced by increasing *N* and *M*. When *N* grows large enough with a fixed *M*, the variations no longer decrease, which indicates that the parameter updating step is stabilized (Supplementary Table S6). Once the parameter updating step is stabilized with a large enough *N*, the *SRMSE*s decrease roughly proportionally with the square root of *M* (Supplementary Table S6). This agrees with the general Monte Carlo method, where the Monte Carlo sampling error decreases proportionally with the square root of the number of random samples (Kroese, et al., 2011). The results also demonstrate that the variations can be well-controlled (less than 5%) with large enough *N* and *M* even when the dimensionality of the parameter space is very large and the *p*-value is extremely small, and the computation time is affordable on a typical desktop computer (Supplementary Table S6).

The above sensitivity analysis provides a general rule for the choice of optimal $N$ and $M$ when applying MCMC-CE in practice, where we can first find the optimal $N$ by examining the variations across different values of $N$ with a fixed $M$. Once the optimal $N$ is determined, we can then increase $M$ to achieve a desired level of accuracy.

## 3. Additional real-world example

Similar to Example 3 in the main text, here we demonstrate the application of MCMC-CE in estimating small $p$-values based on the ratio statistic in differential expression analysis with a microarray gene expression dataset. The dataset used is published in (Golub, et al., 1999) and is available from *bioconductor* as the *golubEsets* package. It contains 38 leukemia patients with 27 acute lymphoblastic leukemia (ALL) patients and 11 acute myeloid leukemia (AML) patients for which the expression levels of 7129 genes are measured. We are interested in testing which genes are differentially expressed between ALL and AML patients. We first perform normalization and variance stabilizing transformation of the data using the package *vsn* (Huber, et al., 2002), and then carry out the differential expression analysis with the same methods as described in Example 3 in the main text (i.e. Brute-force MC, MCMC-CE, Perm0, Perm1 and samr). Except that the numbers of permutations are set as $10^5$ for Perm0 and Perm1, the other settings for those methods are the same as described Example 3 in the main text. Table S10 presents the numbers of significantly differentially expressed genes identified by each method with different FDR threshold values. This example also demonstrates that accurate estimation of small $p$-values is useful for correctly evaluating the genome-wide significance. Of note, the smallest $p$-value estimated by MCMC-CE in this example is $2.80\times10^{-55}$.

**Table S10.** Number of significantly differentially expressed genes identified by each method with different FDR thresholds

| | FDR threshold | | | | | | | |
|---|---|---|---|---|---|---|---|---|
| | **0.0001** | **0.0005** | **0.001** | **0.005** | **0.01** | **0.05** | **0.1** | **0.15** |
| **Brute-force MC** | 355 | 355 | 355 | 422 | 507 | 887 | 1171 | 1403 |
| **MCMC-CE** | 136 | 202 | 240 | 363 | 449 | 887 | 1171 | 1403 |
| **Perm0** | 278 | 278 | 278 | 401 | 542 | 914 | 1235 | 1519 |
| **Perm1** | 0 | 0 | 0 | 272 | 514 | 903 | 1226 | 1512 |
| **samr** | 239 | 239 | 239 | 364 | 530 | 1027 | 1407 | 1857 |

## 4. List of supplementary tables provided as separate files

**Table S1** - "TableS1.Simulation Chi-squared(5).xlsx". This table shows the simulation results with $\chi^2_5$. The relevant texts are in Main Text Section 3.1 and Supplementary Text Section 2.1.

**Table S2** - "TableS2.Simulation Chi-squared(20).xlsx". This table shows the simulation results with $\chi^2_{20}$. The relevant texts are in Main Text Section 3.1 and Supplementary Text Section 2.1.

**Table S3** - "TableS3.Simulation Chi-squared(50).xlsx". This table shows the simulation results with $\chi^2_{50}$. The relevant texts are in Main Text Section 3.1 and Supplementary Text Section 2.1.

**Table S4** - "TableS4.Simulation Chi-squared(100).xlsx". This table shows the simulation results with $\chi^2_{100}$. The relevant texts are in Main Text Section 3.1 and Supplementary Text Section 2.1.

**Table S5** - "TableS5.Simulation Cauchy.xlsx". This table shows the simulation results with the standard Cauchy distribution. The relevant texts are in Main Text Section 3.1 and Supplementary Text Section 2.1.

**Table S6 -** "TableS6.Sensitivity Analysis.pdf". This table presents the sensitivity analysis for combinations of different numbers of random samples used in the parameter updating step (*N*) and in the estimating step (*M*). The relevant texts are in Main Text Section 3.1 and Supplementary Text Section 2.2.

**Table S7** - "TableS7.Gene Set Enrichment Analysis Example.xlsx". This table shows the results of the gene set enrichment analysis example. The relevant texts are in Main Text Section 3.2 Example 1.

**Table S8** - "TableS8.GWAS Example.xlsx". This table shows the results of the GWAS example. The relevant texts are in Main Text Section 3.2 Example 2.

**Table S9** - "TableS9.Differential Expression ACC Data Example.xlsx". This table shows the results of the differential expression analysis with the salivary adenoid cystic carcinoma RNA-Seq data. The relevant texts are in Main Text Section 3.2 Example 3.